\documentclass[11pt,twoside]{article}
\usepackage{./asp2014}

\aspSuppressVolSlug
\resetcounters

\begin{document}

\title{The advanced stages of stellar evolution: impact of mass loss, rotation, and link with B[e] stars}
\author{Cyril Georgy$^1$, Hideyuki Saio$^2$, Sylvia Ekstr\"om$^1$, and Georges Meynet$^1$
\affil{$^1$Geneva Observatory, University of Geneva, Maillettes 51, CH-1290 Sauverny, Switzerland; \email{cyril.georgy@unige.ch}}
\affil{$^2$Astronomical Institute, Graduate School of Science, Tohoku University, Sendai, Japan}}

\paperauthor{Cyril Georgy}{Cyril.Georgy@unige.ch}{0000-0003-2362-4089}{Geneva University}{Geneva Observatory}{Versoix}{}{1290}{Switzerland}
\paperauthor{Hideyuki Saio}{saio@astr.tohoku.ac.jp}{}{Tohoku University}{Astronomical Institute}{Sendai}{}{ 980-857}{Japan}
\paperauthor{Sylvia Ekatr\"om}{Sylvia.Ekstrom@unige.ch}{}{Geneva University}{Geneva Observatory}{Versoix}{}{1290}{Switzerland}
\paperauthor{Georges Meynet}{Georges.Meynet@unige.ch}{}{Geneva University}{Geneva Observatory}{Versoix}{}{1290}{Switzerland}

\begin{abstract}
In this paper, we discuss some consequences of rotation and mass loss on the evolved stages of massive star evolution. The physical reasons of the time evolution of the surface velocity are explained, and then we show how the late-time evolution of massive stars are impacted in combination with the effects of mass loss. The most interesting result is that in some cases, a massive star can have a blue-red-blue evolution, opening the possibility that Blue Supergiants are composed by two distinct populations of stars: one just leaving the main sequence and crossing the HRD for the first time, and the other one evolving back to the blue side of the HRD after a Red Supergiant phase. We discuss a few possible observational tests that can allow to distinguish these two populations, and how supergiant B[e] stars fit in this context.
\end{abstract}

\section{Introduction}
Since its introduction by Conti in 1976, the B[e] phenomenon still remains largely unexplained, despite numerous studies and progresses since. According to \citet{Lamers1998a}, a star is qualified as a B[e] star if the following criteria are fulfilled:
\begin{itemize}
\item Strong Balmer emission lines,
\item Low excitation permitted emission lines of predominantly low ionisation metals in the optical spectrum,
\item Forbidden emission lines of [Fe II] and [O I] in the optical spectrum,
\item A strong near or mid-infrared excess due to hot circumstellar dust.
\end{itemize}
This characteristics imply the presence of a dense circumstellar medium (CSM), density and temperature conditions allowing for the presence of dust, a non spherical CSM, and at least some events of very strong mass losses \citep{Lamers1998a}. \citet{Lamers1998a} have also introduced the following ``classical'' classification for B[e] stars: evolved high mass stars (sgB[e]), intermediate pre-main sequence objects (HAeB[e]), compact planetary nebulae stars (cPNB[e]), symbiotic stars (symB[e]), and unclassified ones (unclB[e]).

In the following, we discuss what are the expected properties of Blue Supergiants (BSGs) in the framework of single star evolution. Particularly, the effects of rotation and mass loss are presented. Finally, we discuss how sgB[e] fit into this context.

\section{Post-Main-Sequence evolution of the surface rotational velocities}

Different factors impact the evolution of the surface velocity of a single (massive) star once it leaves the main sequence and enters the evolved stages: the change in the radius of the star, internal redistribution of angular momentum, and loss of angular momentum due to stellar winds.

\paragraph{Stellar radius} At the end of the main sequence (MS), the star crosses quickly the Hertzsprung-Russell diagram at roughly constant luminosity, due to the so-called ``mirror effect'' \citep[the core contracts, producing an extension of the envelope, \textit{e.g.}][]{Kippenhahn1990a}. For example, a model of a $9\,M_\odot$ star at solar metallicity has a radius of about $7\,R_\odot$ at the end of the MS, and reaches a radius of about $200\,R_\odot$ when the star evolves at the bottom of the red supergiant (RSG) branch. For a $20\,M_\odot$ model, the corresponding values are about $13\,R_\odot$ and $800\,R_\odot$ \citep[models from][]{Ekstrom2012a}. This change in the radius brakes efficiently the stellar surface, by conservation of the angular momentum.

\paragraph{Internal redistribution of angular momentum} In the current framework of the theory of stellar rotation \citep[see \textit{e.g.}][]{Maeder2009a}, various instabilities in stellar interiors can develop and transport angular momentum. In radiative zones, this transport is usually slow, and requires long timescale to act \citep[however, see][]{Beck2012a,Eggenberger2012a,Cantiello2014a}. The quick crossing of the HRD after the MS makes the surface hardly affected by angular momentum transport in radiative zones. As the star reaches the RSG branch, a deep external convective zone develops. Fast redistribution of angular momentum inside the convective zone can slightly affect the rotation velocity of the surface.

\paragraph{Mass loss} Stellar winds are very efficient in removing angular momentum from the stellar surface, particularly when the star becomes cool and exhibits very high mass-loss rates. For stars that will have a bluewards evolution after the RSG phase (second HRD crossing), the mass loss is also efficient in preventing the stellar surface to re-accelerate due to the contraction of the star.

\begin{figure}[ht]
\begin{center}
\includegraphics[width=.45\textwidth]{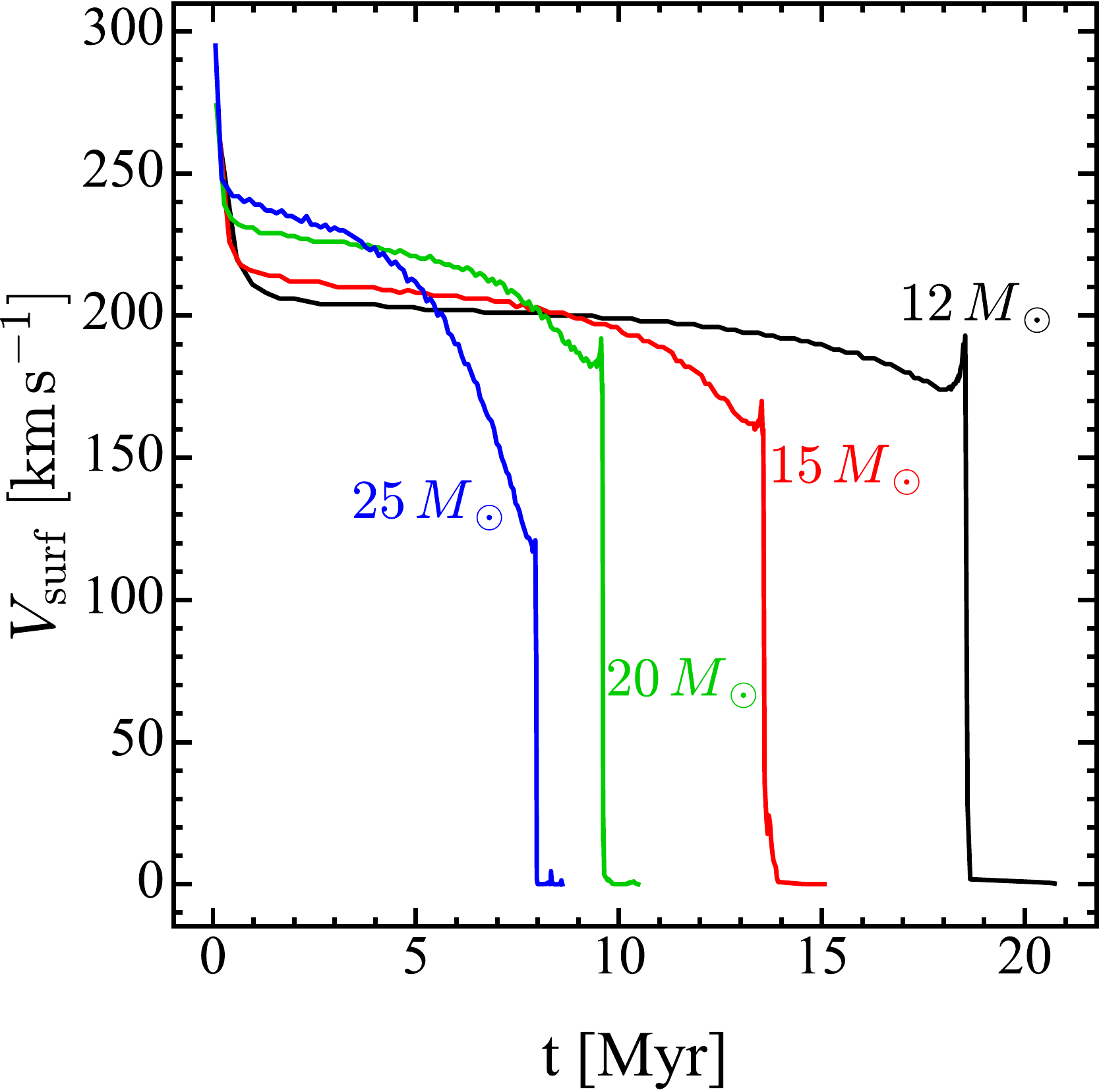}\hfill\includegraphics[width=.5\textwidth]{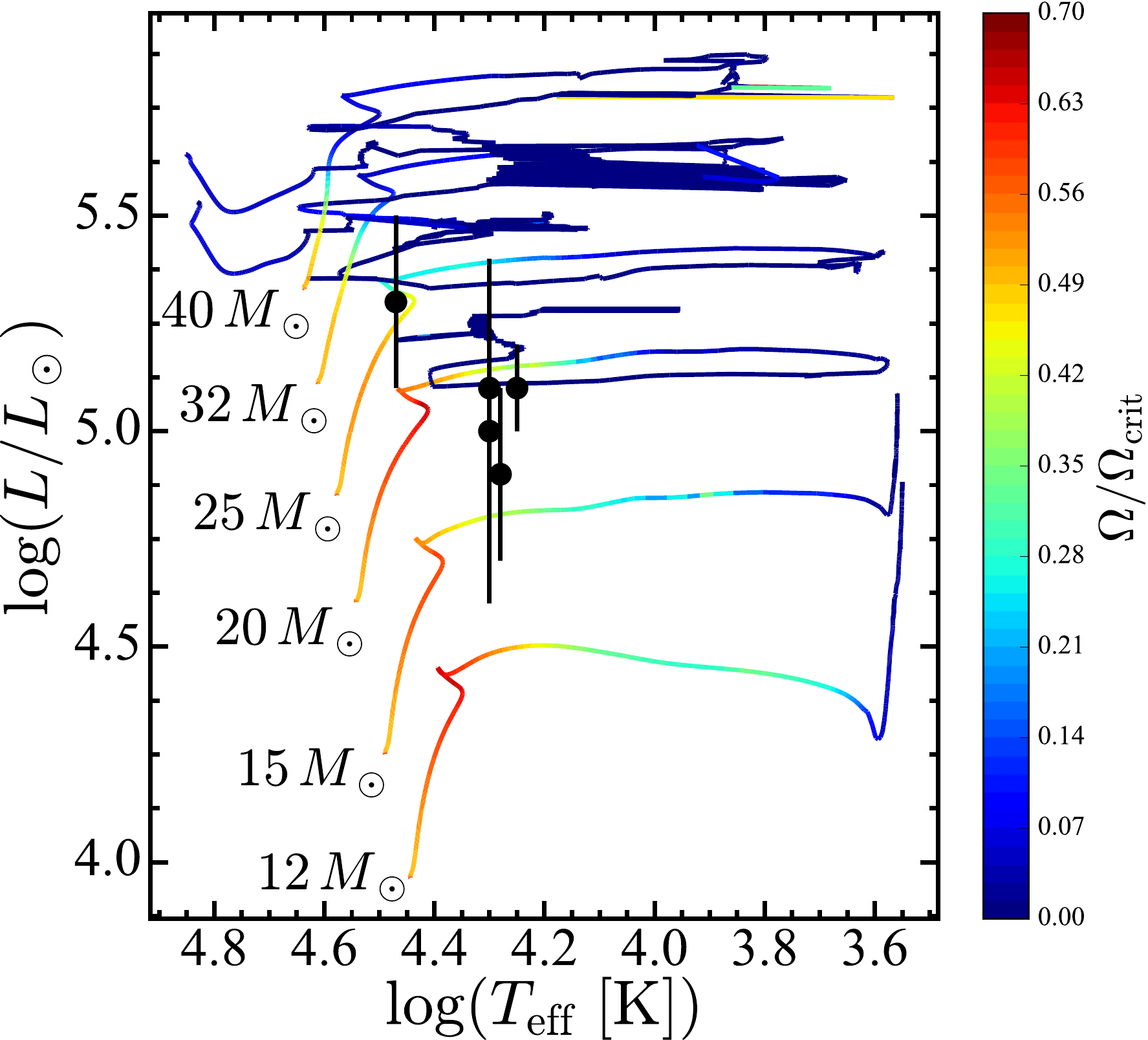}
\end{center}
\caption{\textit{Left panel:} time evolution of the surface equatorial velocity for models at solar metallicity and initial mass of $12$, $15$, $20$, and $25\,M_\odot$. \textit{Right panel:} HRD for models from $12$ to $40\,M_\odot$ at solar metallicity. The surface ratio of the actual angular velocity to the critical one \citep[as defined \textit{e.g.} in][]{Maeder2000a}. Positions of observed sgB[e] stars are indicated \citep[from][note that the colour of the observational points does not correspond to measurements of the velocity of the stars.]{Miroshnichenko2007a}.}
\label{FigVelocity}
\end{figure}

The combined effect on the evolution of the surface velocities is shown in the left panel of Fig.~\ref{FigVelocity}. After a relaxation phase at the very beginning of the MS, during which the initially flat rotation profile inside the star reaches its quasi-stationary regime, producing the quick decrease of the velocity, we see two different behaviours, depending on the initial mass. For the $12\,M_\odot$ model, the equatorial velocity remains almost constant during the MS. The braking due to the progressive extension of the stellar radius is compensated by a flux of angular momentum coming from the central part of the star \citep{Meynet2000a}. For higher mass models, due to more powerful stellar winds, the braking of the surface becomes more and more efficient during the MS, the flux of angular momentum from the interior being insufficient to compensate for the loss due to the winds. In all the cases, the surface velocity drops suddenly immediately after the MS to almost no rotation. Note that the time-evolution of the surface velocity is however quite dependant on the assumed physics for the transport of angular momentum inside the star. In case a strong coupling between the core and the envelope is assumed (as it is supposed to be if internal magnetic fields are accounted for), the model can keep a higher velocity \citep[e.g.][]{Maeder2005a}.

\begin{figure}[ht]
\begin{center}
\includegraphics[width=.6\textwidth]{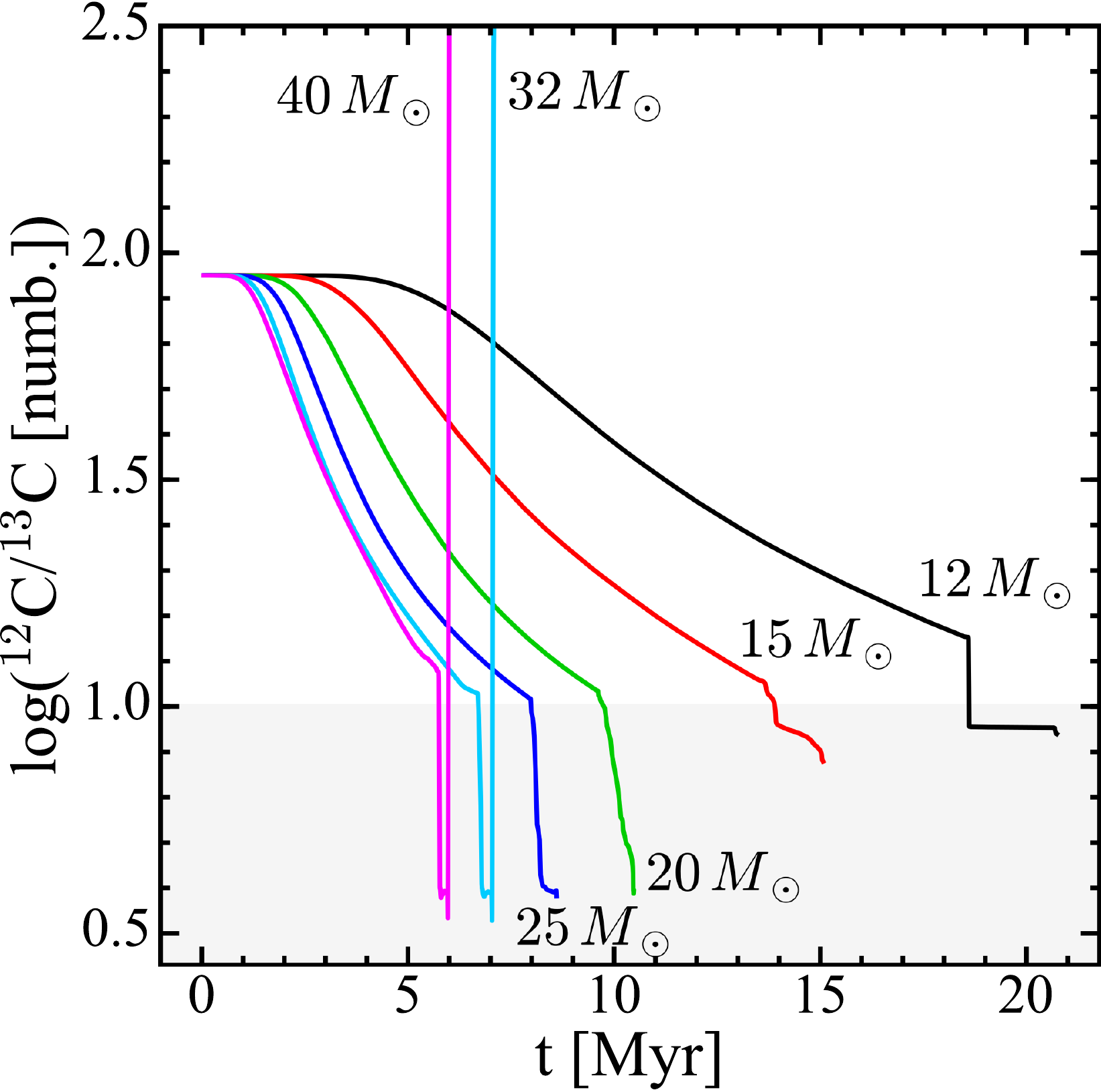}
\end{center}
\caption{Time evolution of the $^{12}\mathrm{C}/^{13}\mathrm{C}$ ratio in different stellar models from \citet{Ekstrom2012a}. The shaded area corresponds to the value of this ratio that are reached only after the MS. The $32$ and $40\,M_\odot$ enter into a WC phase during which $^{12}\text{C}$ increases a lot while $^{13}\text{C}$ decreases.}
\label{FigC12C13}
\end{figure}

In the right panel of Fig.~\ref{FigVelocity} is shown an HRD for rotating massive stars. The colour code indicate the ratio $\Omega/\Omega_\text{crit}$, where $\Omega$ is the actual angular velocity of the stellar surface, and $\Omega_\text{crit}$ the critical angular velocity \citep{Maeder2000a}. Positions of Galactic sgB[e] stars from the sample of \citet{Miroshnichenko2007a} are indicated. The corresponding ratios $\Omega/\Omega_\text{crit}$ on the tracks are about $0.4$. This rather small ratio implies that the expected stellar winds are not very aspheric \citep{Georgy2011a} and if they would be they would rather present enhancement along the polar direction than in the equatorial plane. The models have here surface rotations that are too far away from the critical value to explain an equatorial mechanical mass loss.

\section{Rotational mixing and surface abundances}

One of the important effects of rotation on stellar evolution is that it allows for the development of various instabilities inside the star \citep[e.g.][]{Maeder2009a}. These instabilities trigger transport of chemical species through the star. For most of stars, this mixing will modify the surface abundances of C, N, and O, as well as modifying the isotopic number ratios.

As an illustration, Fig.~\ref{FigC12C13} shows the $^{12}\mathrm{C}/^{13}\mathrm{C}$ ratio as a function of time for several rotating massive stars models between $12$ and $40\,M_\odot$. Due to the action of rotational mixing, the $^{12}\mathrm{C}/^{13}\mathrm{C}$ ratio decreases already during the MS. Indeed, the equilibrium $^{12}\mathrm{C}/^{13}\mathrm{C}$ ratio for H burning through CNO cycle is lower than the initial value, making it change in the burning core. Diffusion of chemical elements due to rotational mixing then slowly changes the surface values accordingly. In the figure, the shaded area corresponds to the values that can be reached only after the MS, and that are a good clue about the evolutionary status of a star. This test can be a powerful tool helping in determining the status of B[e] stars \citep{Kraus2009a,Liermann2010a,Muratore2015a}.

It is worth to mention that the evolution of the surface abundances are however very sensitive to the way internal mixing is implemented into the stellar models. For example, rotational mixing exists in a variety of different flavours, leading to discrepancies in the results obtained with one code, but using different prescriptions \citep{Meynet2013a}, or between different stellar evolution codes \citep[see for example][]{Martins2013a,Chieffi2013a}.

\section{Impact of mass loss during the advanced stages}

\begin{figure}[ht]
\begin{center}
\includegraphics[width=.6\textwidth]{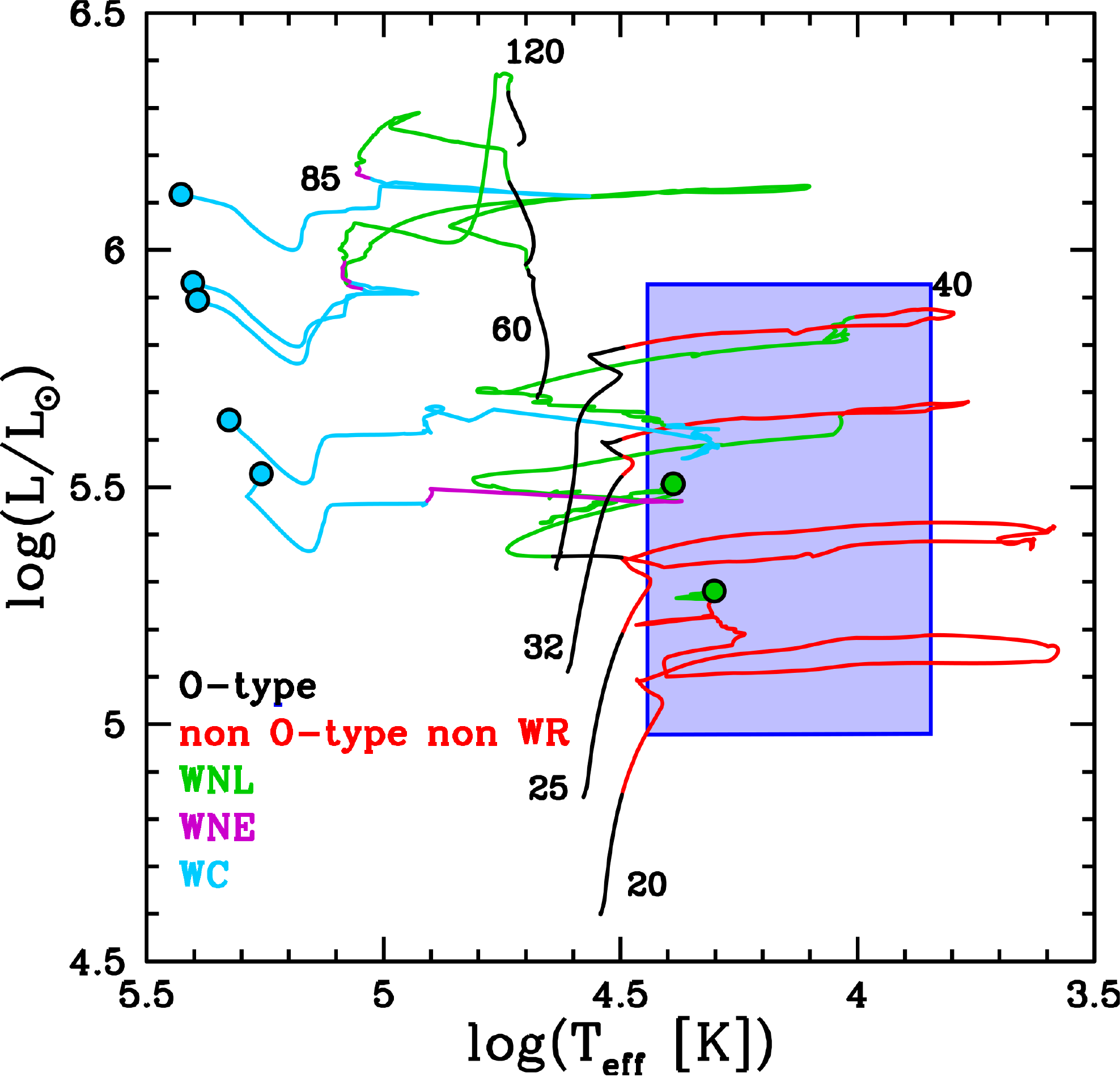}
\end{center}
\caption{HRD for massive star models \citep{Ekstrom2012a}. The colours on the track indicate the stellar type. The blue square indicate the region where BSGs are located, and where pre- and post-RSG phase BSGs are found according to our models. Figure adapted from \citet{Georgy2012b}.}
\label{FigPopBSG}
\end{figure}

Mass loss is a key ingredient of massive star modelling \citep[\textit{e.g.}][]{Conti1976a,Maeder1981a}. However, the current prescriptions used in stellar evolution codes are still uncertain, particularly for the mass-loss rates used during the cool stages of stellar evolution \citep[][see also discussion in \citealt{Georgy2012a}]{vanLoon2005a,Mauron2011a,Beasor2016a}. This uncertainty has huge impact on our knowledge of post-MS evolution of massive stars:
\begin{itemize}
\item By removing a lot of mass during the RSG phase, a strong mass-loss rate makes possible a bluewards evolution of stellar models above $\sim20\,M_\odot$ after the RSG phase \citep{Vanbeveren1998a,Georgy2012a,Georgy2012b,Meynet2015a}. This is a possible solution of the so-called ``Red-Supergiant problem'' \citep{Smartt2009a}, stating that no progenitor of type IIP supernova above about $17\,M_\odot$ have been found while observations (and modelling) shows that there are RSGs (the direct progenitor of the type IIP supernovae) up to about $25\,M_\odot$ \citep[see also][]{Walmswell2012a}.
\item In case the mass-loss rate during the RSG phase is not strong enough to completely remove the hydrogen-rich envelope, it becomes possible for single star to end its evolution at locations in the HRD intermediate between the cool RSGs and the very hot Wolf-Rayet stars, allowing for unexpected supernova progenitors \citep{Georgy2012a,Georgy2012b,Groh2013c,Groh2013a,Groh2013b,Meynet2015a}. For example, it is possible to have stellar evolution tracks ending in the Yellow Supergiants area of the HRD, or in the Luminous Blue Variables one.
\item The blue-red-blue evolution of some of the massive star models makes that the BSG region of the HRD is populated by two different populations of stars: the one that are on their first crossing of the HRD, immediately after the MS, and the one that are evolving back to the blue from the RSG branch (see Fig.~\ref{FigPopBSG}).
\end{itemize}

\section{Distinguishing between both BSG populations}

\begin{figure}[ht]
\begin{center}
\includegraphics[width=.6\textwidth]{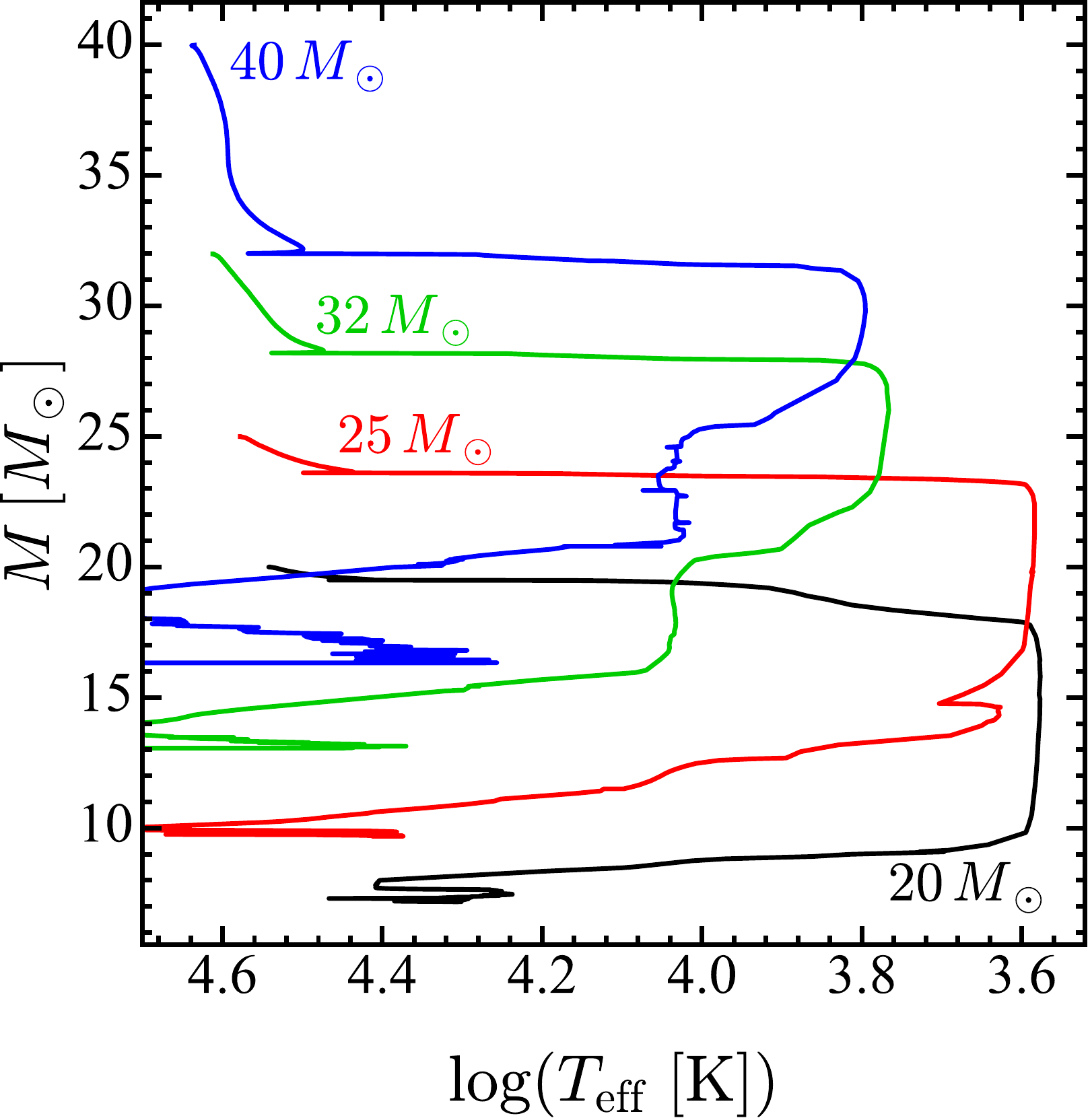}
\end{center}
\caption{Evolution of the total mass of our stellar models \citep{Ekstrom2012a} as a function of the effective temperature. The tracks start at the top, and then evolve down due to mass loss.}
\label{FigTotalMass}
\end{figure}

In the framework of sgB[e] stars, which are also located in the BSG region of the HRD, it is interesting to know whether it is observationally possible to distinguish between a BSG star that is on its first crossing, and another one at the same location of the HRD, but currently on its second crossing. Looking at Fig.~\ref{FigPopBSG}, we see that the models that have a blue-red-blue evolution cross the HRD twice at about the same luminosity. One major difference between a model of BSG on its first or second crossing is the total mass of the star (see Fig.~\ref{FigTotalMass}). Indeed, the mass-loss rates during the RSG phase, even if they are uncertain, are very high (typically about $10^{-6}$ to $10^{-5}\,M_\odot\,\text{yr}^{-1}$). With a RSG phase during a few $100,000\,\text{yr}$ to $1\,\text{Myr}$, a star can lose $5-10\,M_\odot$ during this stage.

Depending on the chemical structure of the star and how internal transport is treated \citep{Georgy2014a}, this huge loss of mass can uncover layers where chemical composition was modified and thus showing signature of CNO-cycle burning. As an illustration, it causes the very quick variations of the $^{12}\text{C}/^{13}\text{C}$ ratio visible in Fig.~\ref{FigC12C13} at the end of the tracks. An other obvious difference between models on the first or second crossing is the surface gravity, considerably lower on the second one (the star evolving roughly at constant luminosity, they have about the same radius on both crossing at a given effective temperature. The mass having considerably decreased makes the gravity to decrease as well).

\begin{figure}[ht]
\begin{center}
\includegraphics[width=.6\textwidth]{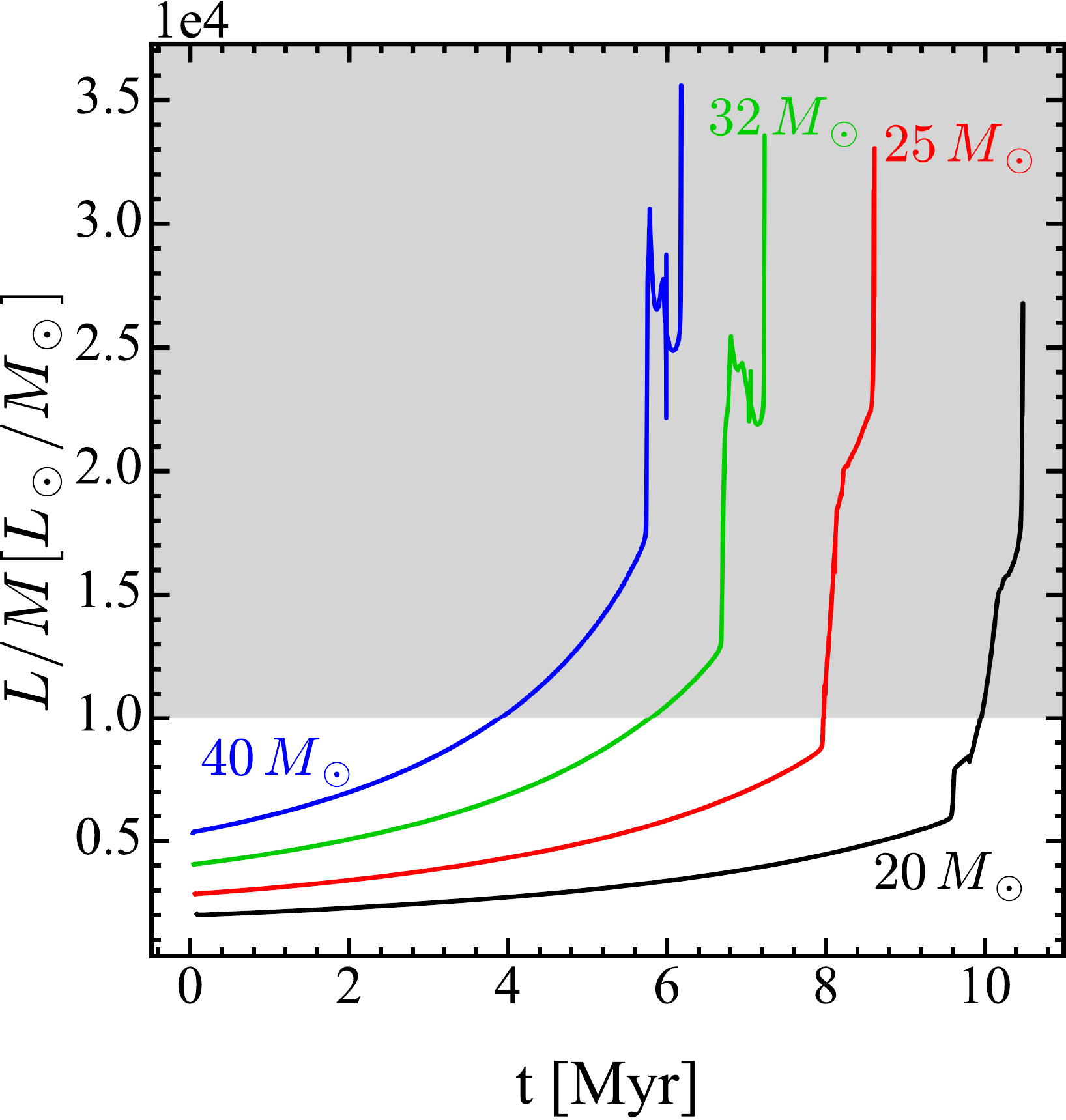}
\end{center}
\caption{Evolution of the luminosity to mass ratio for models of massive star from \citet{Ekstrom2012a}.}
\label{FigML}
\end{figure}

Another direct consequence of the strong mass loss during the RSG phase is that it increases the $L/M$ ratio (see Fig.~\ref{FigML}). This considerably favours the development of strange modes inside the star, which appear when $L/M > 10^4\,L_\odot/M_\odot$ \citep[shaded area in Fig.~\ref{FigML}, see][]{Gautschy1990a,Glatzel1994a,Saio1998a,Saio2013a}. According to our models, BSGs below $~25\,M_\odot$ that exhibit observable pulsation, such as the $\alpha$-Cygni variable, need to have been previously through a RSG phase to evolve back to the blue with a $L/M$ ratio allowing for the apparition of pulsations \citep{Saio2013a}.

Interestingly enough, we have shown that the surface chemical abundances of BSG during the second crossing depend on the way convection is modelled in the stellar evolution code \citep{Georgy2014a}. The observational study of $\alpha$ Cygni variables is thus a powerful tool to constrain models of massive stars.

The existence of $\alpha$ Cygni variables seems to indicate that at least some massive stars in the range $15-25\,M_\odot$ have a blue-red-blue evolution. How often a star evolves back from the RSG to a BSG phase is not known yet. It critically depends on the adopted mass-loss rates during the RSG phase \citep{Meynet2015a}. It has been shown during the last decade that BSGs lie on a tight relation in the flux-weighted gravity vs. bolometric magnitude diagram \citep[the so-called ``Flux-weighted Gravity--Luminosity Relation'' (FGLR), see][]{Kudritzki2003a,Kudritzki2008a,Kudritzki2012a}. In \citet{Meynet2015b}, we have shown that current stellar models are able to well reproduce the observed FGLR, provided that most of the BSGs are on their first crossing of the HRD or that if they are on their second crossing, they have lost little mass in previous phase. This last possibility is however difficult to account in stellar models since mass loss is the key factor allowing the blueward evolution to occur.

\section{Conclusions}

In this paper, we have discussed the properties of BSGs in the perspective of single star models, particularly the evolution of their surface velocities, their mass-loss rates, and the expected existence of two distinct populations of BSGs, on a first and on a second crossing of the HRD. We have shown how both these populations can be distinguished observationally on the basis of their pulsational properties.

In case (some of the) sgB[e] stars arise from single star evolution, the strong mass-loss mechanism of these object is still unknown. Moreover, it is hard to explain the asymmetries in the CSM of these objects, since our models predict small rotation rates for BSGs at the observed position of sgB[e] stars in the HRD. Another possibility is a binary origin of these object, that deserves further investigations \citep[\textit{e.g.}][]{deWit2014a}.

\clearpage 

\acknowledgements{CG, SE, and GM acknowledge support from the Swiss National Science Foundation (project number 200020-160119).}

\bibliographystyle{asp2014}
\bibliography{georgy_bep2016}  

\end{document}